\documentclass[12pt,onecolumn]{IEEEtran}

\usepackage{amsmath,amssymb,euscript ,yfonts,psfrag,latexsym,dsfont,graphicx,bbm,color,amstext,wasysym,subfig,parskip}
\usepackage{soul}
\graphicspath{{./},{./figures/}}

\begin{document}
\newcommand{\be}{ \begin{equation} }
\newcommand{\ee}{ \end{equation} }
\newcommand{\cM}{{\cal M}}

\title{Graph Curvature and the Robustness of Cancer Networks}

\author{Allen Tannenbaum, Chris Sander, Romeil Sandhu, Liangjia Zhu, Ivan Kolesov, Eduard Reznik, Yasin Senbabaoglu, Tryphon Georgiou\thanks{A. Tannenbaum, R. Sandhu, L. Zhu, and I. Kolesov are with the Department
of Computer Science, Stony Brook University: (emails: \{allen.tannenbaum,romeil.sandhu,liangjia.zhu, ivan.kolesov\}@stonybrook.edu). C. Sander, E. Reznik, and Y. Senbabaoglu are with the Computational Biology Program of Memorial Sloan Kettering: (\{emails: sander, eznike, yasin\}@cbio.mskcc.org). T. Georgiou is with the Department of Electrical and Computer Engineering, University of Minnesota: (email: tryphon@umn.edu) }}

\maketitle

\begin{abstract}

The importance of studying properties of networks is manifest in numerous fields ranging from biology, physics, chemistry, medicine, neuroscience, and engineering. The functionality of networks with regard to performance, throughput, reliability and robustness is strongly linked to the underlying geometric and topological properties of the given network, which is the focus of this paper, especially as applied to certain biological networks.
The fundamental mathematical abstraction of a network as a weighted graph brings to bear the tools of graph theory--a highly developed subject of mathematical research. But more importantly, recently proposed geometric notions of curvature on very general metric measure spaces allow us to utilize a whole new set of tools and ideas that help quantify functionality and robustness of networks modelled as graphs.
In particular,
robustness is closely connected to network entropy which, in turn, is very closely related to curvature.  There are a number of alternative notions of discrete curvature that are compatible with the classical Riemannian definition, each having its own
advantages and disadvantages,
and are relevant to networks of interest. We will concentrate on the role of curvature for certain key cancer networks in order to quantitatively indicate their apparent functional robustness relative to their normal counterparts.
\end{abstract}

\section{Introduction}\label{sec:Introduction}

The growing importance of studying complex networks has been documented in a huge and growing literature, and has even been referred to as the field of {\em network science} \cite{Barabasi}. We are interested in the deep and intimate relationship between network functionality in terms of robustness, on the one hand, and topological and geometric properties such as curvature on the other. Following \cite{dem}, we will give a formal definition of ``robustness'' in terms of the rate function from large deviations theory. This definition captures the essential property of complex networks of being able to adapt
to dynamic changes (modelled as random perturbations), while maintaining functionality \cite{alon}. One of the key ideas is based on the positive correlation between an increase of curvature and network functional robustness, as we will argue in Section~\ref{sec:curve_robust}. As we will explain, the necessary geometric notions are quite easy to compute in terms of the graph Laplacian or the Earth Mover's Distance and, thus, one can assess and quantify robustness of networks quite efficiently. This relationship is particularly compelling because, on one hand, curvature is an equilibrium property of the system (as is the entropy) while, on the other, robustness is a dynamic (non-equilibrium) property. Nevertheless this correlative relationship holds.

We will also see that network curvature is closely related to network entropy. In fact, one of the ways of generalizing curvature to rather general metric measure spaces is to exploit the convexity properties of the entropy along geodesic paths defined on the associated space of probability measures with an induced Riemannian structure \cite{Romania,LV,McC97}. In the evolutionary biology literature, it has been noted that entropy is closely related to network topology; see \cite{dem}. In the latter work, the authors argue (based on Darwinian principles) that entropy is a selective criterion that may account for the robustness and heterogeneity of both man-made and biological networks. As we have just mentioned, in this present work, we will argue that curvature from \textbf{\emph{network geometry}} is also strongly related to functional robustness.

Much of the work in this paper is motivated by the problem of seeking basic principles and commonalities across many types of complex networks in addition to taking the data-driven point of view \cite{barabasi}. Indeed, search engines and cell biology have a number of things in common; they need to process incomplete and noisy data and rely on input that is stochastic in nature. Search engines rely on empirical distributions that are based on very limited finite sampling, often of poorly understood features. Cell biological cycles are characterized by empirical distributions that are driven by poorly understood environmental cues and intracellular checkpoints.
This motivates problems that include the characterization of robustness, reliability, and possible uncertainty principles as they pertain to information transfer across networks.

In this note, the central object of study will be cancer networks, and the possibility of using graph curvature information to identify key hubs and possible targets of interest in a quantitative manner.
Indeed, for biological networks in connection with various types of cancer, it appears that entropy is a key network property in revealing ``cancer hallmarks'' \cite{West}; thus, generalizations of curvature that are based on entropy \cite{LV} would seem most appropriate in this context. We will include some background material to make our presentation relatively self-contained. Thus this note will have a tutorial flavor as well. The present work is an extended version of \cite{cancer_arxiv}.

\section{Background on curvature}

Various networks may be regarded as geodesic metric spaces, say via the hop metric for communication networks \cite{Kennedy}. Since the object of interest for a cancer network will be a weighted graph (see Section~\ref{sec:cancer.net}), we will consider notions of curvature that best fit this mathematical model, and can lead to interesting new quantitative biological insights. Accordingly, we will first sketch some material on curvature, before moving on to the proposed notions for networks modelled as graphs.

\subsection{Wasserstein distance}

We begin by recording the basic definition of the $L^p$-Wasserstein distance from optimal transport theory that we will need below. For full details about the Monge-Kantorovich (optimal mass transport) problem and the associated Wasserstein distance, we refer the interested reader to \cite{Evans1989,Rachev,Villani2,Villani3}. Let $X$ be a metric measure space, equipped with distance $d$. Let $\mu_i, \; i=1,2$, be two measures with the same total mass and finite $p$-th moment. A \emph{coupling} between $\mu_1$ and $\mu_2$ is a measure $\mu$ on $X \times X$ such that $$\int_y d\mu(x,y) = d\mu_1(x), \;\;  \int_x d\mu(x,y) = d\mu_2(y).$$ In other words, the marginals of $\mu$ are $\mu_1$ and $\mu_2.$ Let $\Pi(\mu_1, \mu_2)$ be the set of couplings between $\mu_1$ and $\mu_2$.

We define the $L^p$ \emph{Wasserstein distance} as
$$W_p (\mu_1, \mu_2):= \left ( \inf_{\mu \in \Pi(\mu_1, \mu_2)} \int \int d(x,y)^p d\mu(x,y) \right )^{1/p}.$$
In this paper, we will only consider the cases $p=1,2$. For $p=1$, the Wasserstein distance is sometimes called the ``Kantorovich-Rubinstein distance'' or the Earth Mover's distance (EMD).

\subsection{Generalities on Ricci curvature}

In order to motivate generalized notions of Ricci curvature suitable for complex networks, we will begin with an elementary treatment of curvature following \cite{DoCarmo, Villani2,Villani3}. For $\cM$ an $n$-dimensional Riemannian manifold, $x \in \cM$, let $T_x \cM$ denote the tangent space at $x$, and $u_1, u_2 \in
T_x \cM$ orthonormal vectors.
Then for geodesics $\gamma_i(t) :=  \exp(t u_i)$, $i=1,2$, the {\em sectional curvature} $K(u_1,u_2)$ measures the deviation of geodesics relative to Euclidean
geometry, i.e., \begin{equation} \label{geo:ricci} d(\gamma_1(t), \gamma_2(t)) = \sqrt{2} t (1 - \frac{K(u_1, u_2)}{12} t^2 + O(t^4)).\end{equation}
The Ricci curvature is the average sectional curvature. Namely, given a (unit) vector $u \in T_x \cM$, we complete it to an orthonormal basis,
$u, u_2, \ldots, u_n$. Then the {\em Ricci curvature} is defined by $Ric(u):= \frac{1}{n-1}\sum_{i=2}^n K(u, u_i).$ (There are several different scaling factors used in the literature. We have followed \cite{DoCarmo}.) It may be extended to a quadratic form,
giving the so-called {\em Ricci curvature} tensor. Ricci curvature is strongly related to the Laplace-Beltrami operator. Indeed, in geodesic normal coordinates we have
that $$R_{ij} = -3/2 \Delta g_{ij},$$ where $g_{ij}$ denotes the metric tensor on $\cM$.

\begin{figure}[htb]\begin{center}
\includegraphics[totalheight=8cm]{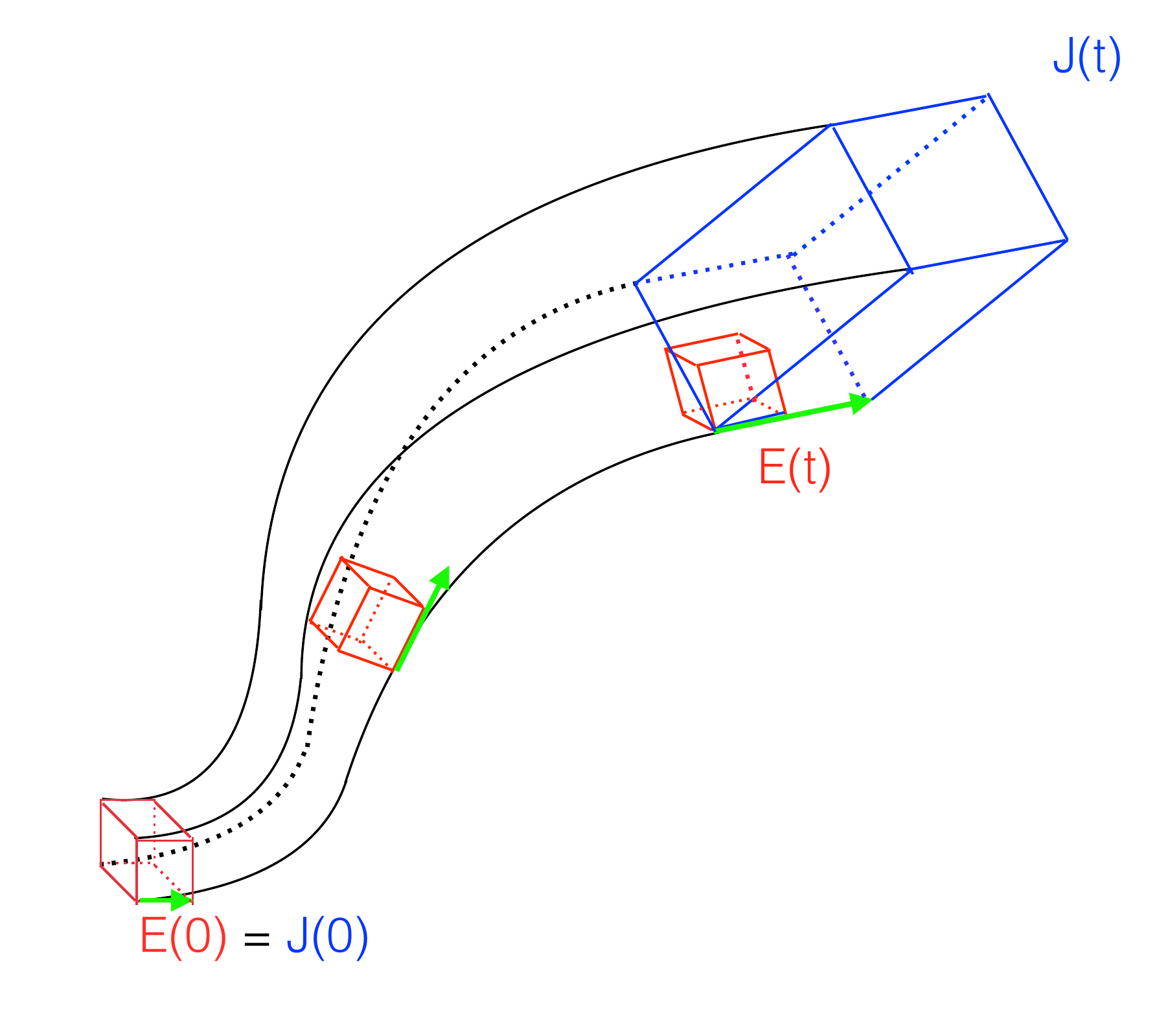}
\caption{Explanation of curvature in terms of Jacobian.}\label{fig:jacobian}\end{center}
\end{figure}

There is another way to understand curvature: it gives a related measure of the spreading of geodesics in the following manner \cite{DoCarmo}.
Let $\gamma$ denote a geodesic and $\gamma_s$ a smooth one parameter family of geodesics with $\gamma_0 =\gamma.$ Then a {\em Jacobi field}
may be defined as
$$J(t) = \frac{d\gamma_s(t)}{ds} \rvert_{s=0}.$$ It may be regarded as an infinitesimal deformation of the given geodesic. Then it is standard that $J(t)$
(essentially the Jacobian of the exponential map) satisfies the Jacobi equation:
\begin{equation} \label{eq:Jacobi} \frac{D^2}{dt^2}J(t) + R(J(t), \dot{\gamma}(t))\dot{\gamma}(t) =0,\end{equation}  where $\frac{D}{dt}$ denotes covariant
derivative, and $R$ is the Riemann curvature tensor;
see Figure~\ref{fig:jacobian}.

We want extend these notions to discrete graphs and networks. For discrete spaces corresponding to networks modeled as graphs, ordinary notions such as
differentiability needed to define Ricci curvature as in the previous section do not make sense. There is however a very nice argument due to
Villani \cite{Villani3} that indicates a possible way to getting around such difficulties via two approaches to convexity.
More precisely, let $f: {\mathbb R}^n \rightarrow {\mathbb
R}.$
Then if $f$ is $C^2$, convexity may be characterized as $$\nabla^2 f(x) \ge 0$$ for all $x$. Villani calls this an {\em analytic} definition of convexity (as the
usual definition of Ricci given above). On the other hand, one can also define convexity in a {\em synthetic} manner via the property that
$$f((1-t)x+ ty) \le (1-t)f(x) + t f(y),$$ for all $x,y \in {\mathbb R}^n,$ and $t \in [0,1].$ In the latter case, no differentiability is necessary.

Following \cite{LV, McC97}, one may define a synthetic notion of Ricci curvature in terms of so-called {\em displacement convexity} inherited from the Wasserstein geometry on probability measures. In \cite{LV}, this is done for measured length spaces, that is, metric measure spaces in which the distance between two points equals the infimum of the lengths of curves joining the points. For discrete spaces (such as those arising in network theory), there are several possibilities that we will compare, especially those from \cite{Romania, Gromov, Ollivier}.
There have been several other approaches as well to defining Ricci curvature; see \cite{Mass,Zhou} and
the references therein.

To conclude, let us just give one elegant result \cite{Villani2}. Assume that $\cM$ is a Riemannian manifold, and let ${\cal P}(\cM)$ denote the space of probability measures on $\cM$ equipped with the Wasserstein 2-metric.  Let $\mu_t$ denote a geodesic curve on ${\cal P}(\cM)$. Then $Ric \ge 0$ if and only if the curve $t \rightarrow \int \mu_t \log \mu_t$ is convex (in $t$). Thus lower bounds on the Ricci curvature are closely connected to entropy and geodesics on spaces of probability measures.

\subsection{Curvature and robustness} \label{sec:curve_robust}

As previously remarked, there have been a number of approaches (see
\cite{Romania,Mass,Zhou,Ollivier} and the references therein) to extending the notion of Ricci curvature to more general metric measure spaces. At this point, the exact relationship of one approach as compared to another is unclear. Roughly, the techniques fall into two categories:
the first generalizing the weak $k$-convexity of the entropy functional on the Wasserstein space of probability measures as in \cite{LV, McC97, Romania}, and the
second directly working with Markov chains to define the generalization on networks as in \cite{Ollivier,Mass,Zhou}. Finally there is a notion of ``hyperbolicity'' due to Gromov \cite{Gromov} based on the ``thinness'' or ``fatness'' of triangles compared to the Euclidean case, and more generally a certain four-point criterion. Depending upon the application, each approach seems to
be useful, and so we will explicate key aspects and their possible applications to various network problems. In particular, we describe in some detail the formulations of curvature in
\cite{LV,Ollivier,Oll_markov} because of connections to notions of metric entropy.

We begin with a characterization given in Lott and Villani \cite{LV}. Let $(X,d,m)$ denote a geodesic space, and set
\begin{eqnarray}
{\cal P}(X,d,m)&:=& \{ \mu \ge 0: \int_X \mu \, dm =1\},\\
{\cal P}^*(X,d,m) &:=& \{ \mu \in {\cal P}(X,d,m): \lim_{\epsilon \searrow 0} \int_{\mu \ge \epsilon} \mu \log \mu \, dm < \infty\}.\end{eqnarray}
We define \be \label{Bent} H(\mu) := \lim_{\epsilon \searrow 0} \int_{\mu \ge \epsilon} \mu \log \mu \, dm,
\mbox{ for } \mu \in {\cal P}^*(X,d,m),
\ee
which is the negative of the {\em Boltzmann entropy}
$S(\mu):=-H(\mu)$; note that the concavity of $S$ is equivalent to the convexity of $H$.
Then we say that $X$ has {\em Ricci curvature bounded from below by $k$} if
for every $\mu_0, \mu_1 \in {\cal P}(X),$ there exists a constant speed geodesic $\mu_t$ with respect to the Wasserstein 2-metric connecting $\mu_0$ and
$\mu_1$ such that
\be \label{eq:entropy.curvature} S(\mu_t) \ge t S(\mu_0) + (1-t) S(\mu_1)+\frac{kt(1-t)}{2} W(\mu_0, \mu_1)^2, \quad 0\le t\le 1.\ee

This indicates the \textbf{\emph{positive correlation}} of entropy and curvature that we will express as
\be \label{ricci.entropy} \Delta S \times \Delta Ric \ge 0.\ee  We will describe appropriate notions of Ricci curvature and entropy on graphs below.
We just note here that changes in {\em robustness}, i.e., the ability of a system to functionally adapt to changes in the environment (denoted as
$\Delta R$) is also positively correlated with entropy via the Fluctuation Theorem \cite{fluctuation,demo}, and thus with network curvature:
\be \label{ricci.robustness} \Delta R \times \Delta Ric \ge 0.\ee
Since the curvature is very easy to compute for a network, this may be used as an alternative way of expressing functional robustness.

Based on the work of \cite{West,dem}, it seems that many normal cellular protein interaction networks have lower entropy than their cancerous analogues. Functionally this makes sense: since oncoproteins are many times disordered (have a larger number of degrees of freedom), they are able to respond better to changes in the
cellular environment. \textbf{\emph{The notion of curvature now allows to precisely quantify this adaptive ability.
}} In Section~\ref{sec:cancer.net}, we will apply curvature to certain cancer networks to test the idea robustness expressed in a geometric manner.

\subsection{Remarks on the Fluctuation Theorem}

The form of the Fluctuation Theorem that we use was formulated in \cite{fluctuation,demo}. The idea is very simple and comes from the theory of large deviations
\cite{varadhan}. One considers random fluctuations of a given network that result in perturbations of some observable. We let $p_\epsilon (t)$ denote the
probability that the mean deviates by more than $\epsilon$ from the original (unperturbed) value at time $t$. Since $p_\epsilon(t) \rightarrow 0$ as $t \rightarrow \infty,$ we want
to measure its relative rate, that is, we set
$$R := \lim_{t \rightarrow \infty} (-\frac{1}{t} \log p_\epsilon(t)).$$ Therefore, large $R$ means not much deviation and small $R$ large deviations. In
thermodynamics it is well-known that entropy and rate functions from large deviations are very closely related. The Fluctuation Theorem is an expression of
this fact for networks, and may be expressed as
\be \Delta S \times \Delta R \ge 0.\ee

The Fluctuation Theorem has consequences for just about any type of network: biological, communication, social, or neural. In rough terms, it means that the ability of a
network to maintain its functionality in the face of perturbations (internal or external), can be quantified by the correlation
of activities of various elements that comprise the network. In the standard statement, this correlation is given via entropy. We have formulated this now
in terms of curvature.

\section{Ollivier-Ricci curvature}

\begin{figure}[htb]\begin{center}
\includegraphics[totalheight=4cm]{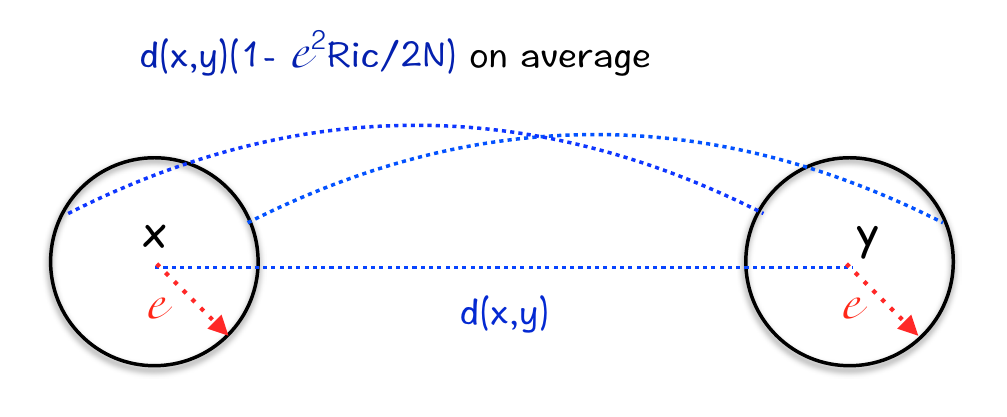}
\caption{Positive curvature is detected when distance from one ball to another is smaller than those of their centers on the average.} \label{fig:balls}\end{center}
\end{figure}

For the weighted graphs used to model the cancer networks of interest in this paper, we will employ a neat notion of Ricci curvature
due to \cite{Ollivier,Oll_markov}. The approach was inspired from several different directions from properties of Ricci curvature in the continuous case. Referring to Figure~\ref{fig:balls}, the idea is that for two very close points $x$ and $y$ with tangent vectors $w$ and $w',$ in which $w'$ is obtained by a parallel transport of $w$, the two geodesics will get closer if the curvature is positive. This is reflected in the fact that the distance between two small (geodesic balls) is less than the distance of their centers. Ricci curvature along direction $xy$ reflects this, averaged on all directions $w$ at $x.$ Similar considerations apply to negative and zero curvature \cite{von}.

More formally, we define for $(X,d)$ a metric space equipped with a family of probability measures $\{\mu_x : x \in
X\},$ the {\em Olliver-Ricci curvature} $\kappa(x,y)$ along the geodesic connecting $x$ and $y$ via
\begin{equation} \label{OR} W_1(\mu_x,\mu_y) = (1-\kappa(x,y))d(x,y),\end{equation}
where, as noted earlier, $W_1$ denotes the Earth Mover's Distance (Wasserstein 1-metric), and $d$ the geodesic distance on the graph.
For the case of weighted graphs of greatest interest in networks, we put
\begin{eqnarray*} d_x &=& \sum_y w_{xy}\\
\mu_x(y)&:=& \frac{w_{xy}}{d_x} ,\end{eqnarray*}
the sum taken over all neighbors of $x$ where $w_{xy}$ denotes the weight of an edge connecting $x$ and $y$ (it is taken as zero if there is no connecting
edge between $x$ and $y$). The measure $\mu_x$ may be regarded as the distribution of a one-step random walk starting from $x$. As is argued in
\cite{Ollivier}, this definition is more inspired from an approach such as that given via equation~(\ref{geo:ricci}). An advantage of this, is that it is
readily computable since the Earth Mover's Metric may be computed via linear programming \cite{Villani3}.

Note that if we define the Laplacian operator via
$$\Delta f(x) = f(x) -\sum_y f(y)\mu_x(y), \; \mbox{$f$ real-valued function},$$ this coincides with the usual normalized graph Laplacian operator
\cite{Jost}. It is also interesting to note in this connection that if $k \le \kappa(x,y)$ is a lower bound for the Ricci curvature, then
the eigenvalues of $\Delta$ may be bounded as $k \le \lambda_2 \le \ldots \lambda_{N}\le 2-k$; see \cite{Jost} for the exact statement. (The first eigenvalue $\lambda_1=0.$) This
relationship is very important since $2-\lambda_{N}$ measures the deviation of the graph from being bipartite, that is a graph whose vertices can be
divided into two disjoint sets $U$ and $V$ such that every edge connects a vertex in $U$ to one in $V$. Such ideas appear in resource allocation in certain networks.

Lastly, while this work demonstrates the usage of curvature to quantify robustness of networks (and in particular cancer networks), altering the network properties to uncover potential drug targets provides an exciting avenue.  With this in mind, one may consider the corresponding Ricci flow:
\begin{equation} \label{eq:dis.ricci} \frac{d}{dt}d(x,y) = -\kappa(x,y) d(x,y). \end{equation} Not much is known about this flow, but the idea would be, while keeping the same topology, to change the graph weights, or the network of links among the nodes, in such a way as to uniformize the curvature $\kappa$. As argued in \cite{Jonck}, this could have the effect (in case of certain wireless networks) of removing some of the overloaded queues, and thus may have important implications for cancer networks as well. Understanding discrete analogues of Ricci flow in this connection will be considered as a future research topic in this connection. Finally, we should note that the Ollivier-Ricci curvature gives the value of the ordinary Ricci curvature on a Riemannian manifold after scaling.

\subsection{Examples of fluctuation and curvature}

\textbf{Mean-Reverting Process:}  It is very informative to consider the relationship of the Ollivier-Ricci curvature and robustness via a simple example \cite{Ollivier,Oll_markov}.
Consider the Ornstein-Uhlenbeck process
\begin{equation} \label{eq:OU}
dX_t = -\alpha X_t dt + \sigma dW_t, \;\;  X(0) =x_0, \end{equation}
where $W$ is Brownian motion (Wiener process), and we take $x_0$ to be deterministic. We treat the 1-dimensional case for simplicity. Everything goes through in higher dimensions as well. The corresponding Fokker-Planck equation is
\begin{equation} \label{eq:fokker}
\frac{\partial p}{\partial t} = \alpha \frac{\partial xp}{\partial x} + \frac{\sigma^2}{2}\frac{\partial^2 p}{\partial x^2},\end{equation}
where $p=p(x,t|x_0,0)$ is the transition probability of the underlying Markov process. One may show that $p(x,t|x_0,0)$  is a Gaussian process with mean and variance given by \cite{Gardiner}:
$$\langle X(t) \rangle = x_0 e^{-\alpha t},\quad \mbox{var}{X(t)}= \frac{\sigma^2}{2 \alpha} (1- e^{-2\alpha t}).$$

We see that we get transition probabilities of mean $x_0 e^{-\alpha t}$ and variance independent of $x_0$. Since all the transitions $p(x,t|x_0,0)$ have the same variance (and are Gaussian) the 1-Wasserstein distance \cite{Givens}
$$W_1(p(x, t|x_0,0), p(x,\delta t|x_1,0))=
|x_0 e^{-\alpha  t} - x_1 e^{-\alpha t}|.$$ Finally,
\begin{equation} \label{ricci_ou} \kappa(x_0,x_1) =1-  \frac{W_1(p(x, t|x_0,0), p(x, t|x_1,0))}{|x_0-x_1|}= 1- e^{-\alpha t}.\end{equation}

\begin{figure}[t]
\mbox{ \centering
    \parbox{.45\columnwidth}{\centering
      \includegraphics[scale=.5]{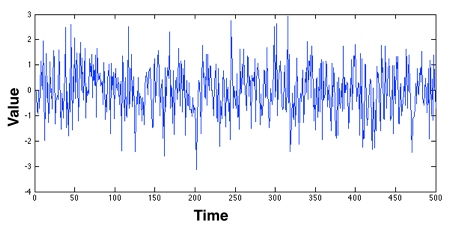}
      (a) }
    \parbox{.45\columnwidth}{\centering
       \includegraphics[scale=.5]{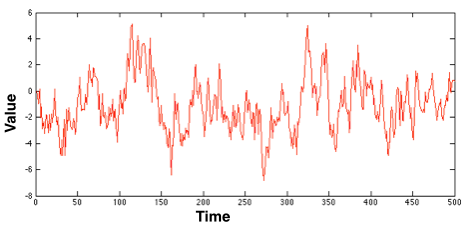}\\
       (b)}}
\\\caption{
(a) OU process with parameters: $\alpha= 1.0$, $\sigma$= 1, $x(0)=1$.  Ollivier-Ricci Curvature Computed at $t = 1$: $\kappa(x,y)= 0.6321$ (b) OU process with parameters: $\alpha= 0.1$, $\sigma$= 1, $x(0)=1$. Ollivier-Ricci Curvature Computed at $t = 1$: $\kappa(x,y)= 0.0952$. }
\label{fig:OUProc}
\end{figure}

\begin{figure}[t]
\mbox{ \centering
    \parbox{.45\columnwidth}{\centering
      \includegraphics[scale=.185]{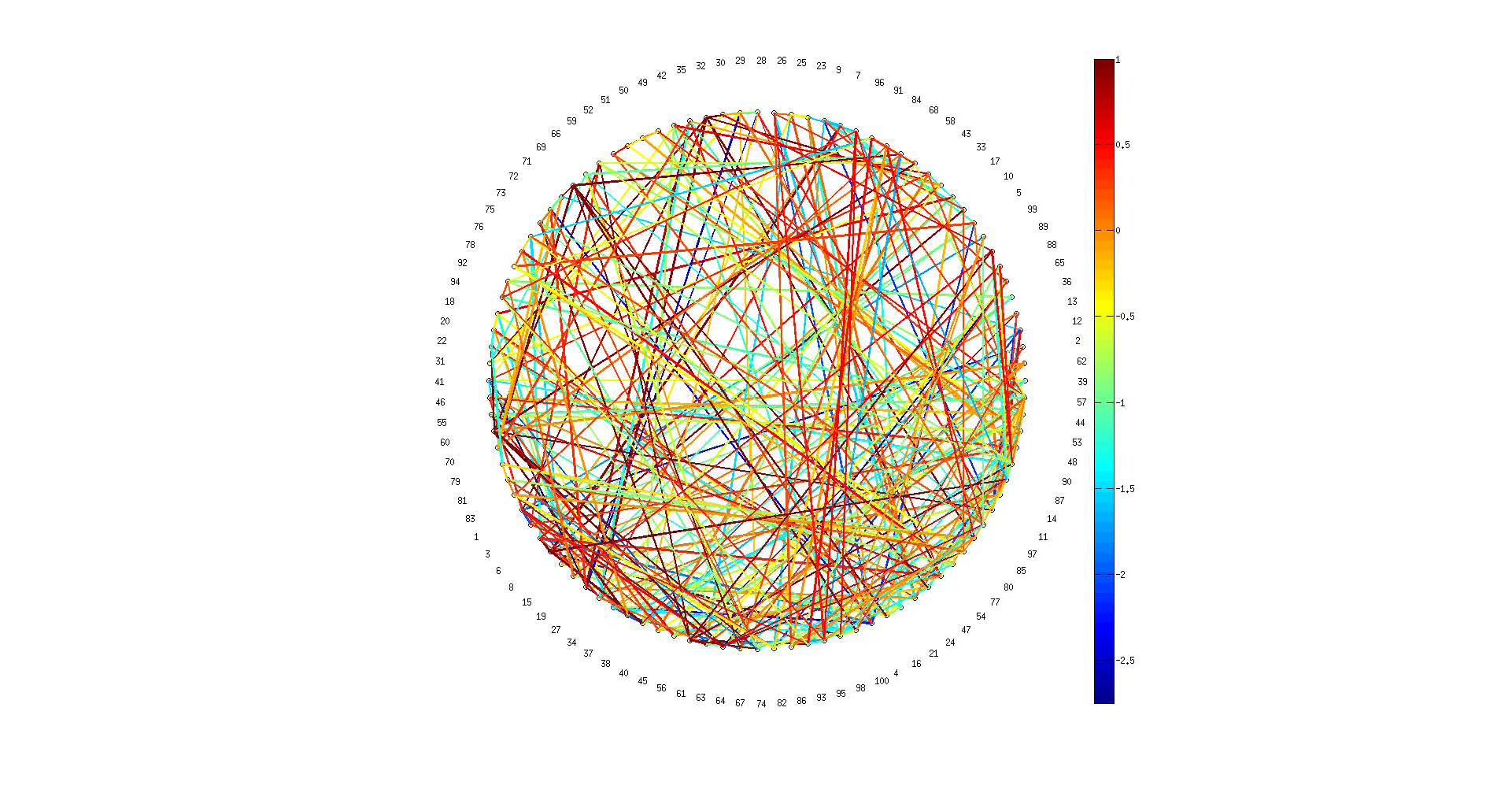}
      (a)}
    \parbox{.45\columnwidth}{\centering
       \includegraphics[scale=.18]{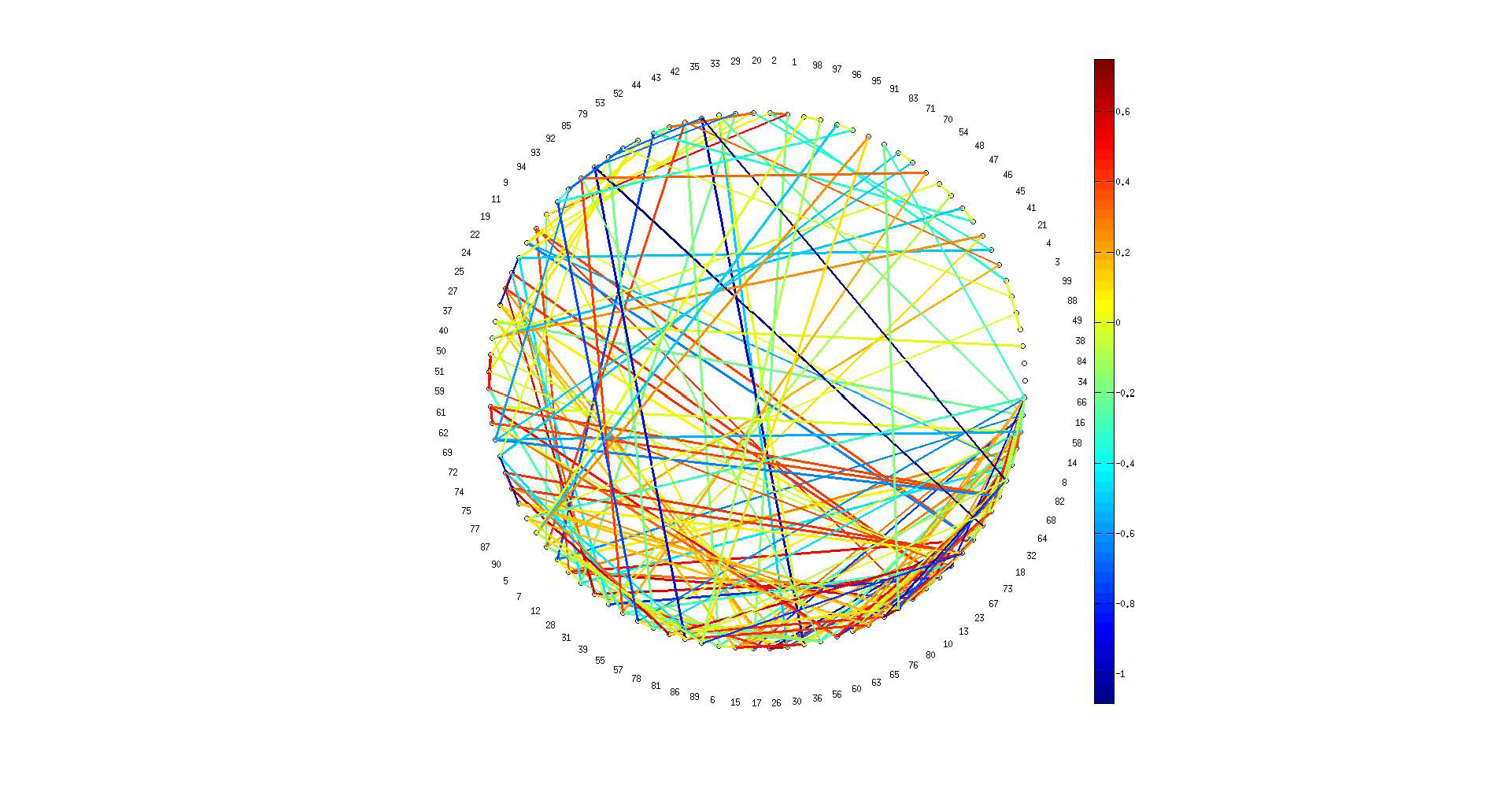}\\
       (b)}}
\\\caption{
(a) Olliver-Ricci curvature of random graph: darker colors represent more negative curvature  (b) Olliver-Ricci curvature of transcription network of {\em E. Coli}: darker colors represent more negative curvature.}
\label{fig:ricci_graph}
\end{figure}

Equation~(\ref{ricci_ou}) illustrates the connection of fluctuation in a very simple explicit manner. Larger $\alpha$ corresponds to larger curvature $\kappa$ and this corresponds to how quickly the systems returns to equilibrium, that is to the mean going to 0. This can be seen in Figure \ref{fig:OUProc} for two given processes with differing $\alpha$.
\\
\\
\textbf{Random vs Biological Graphs:} The second example involves the computation of the Ollivier-Ricci curvature for a random graph as compared to that of a widely available biological network (transcription network of {\em E. Coli} taken from \cite{data_bases}).  Figure~\ref{fig:ricci_graph} shows the results for this computation of both networks.
It has already been argued by Uri Alon \cite{alon} that this transcription network exhibits a great degree of robustness and this is shown here. We also note that the random network has
average Ricci curvature of about $-0.352085 $ with 63\% of the edges being negative, while for {\em E. Coli} we have average curvature of $-0.0222$ with 50\% of the edges being negative.

\subsection{Scalar curvature}
Until now, we have considered Ricci curvature (in the Ollivier sense), which is defined between \emph{any} two vertices on a graph.  One may consider other forms of curvature such as scalar curvature to further illuminate certain geometric network properties. Specifically, scalar curvature represents the amount by which the volume of a geodesic ball in a curved Riemannian manifold deviates from that of the standard ball in Euclidean space \cite{DoCarmo}. Given our previous definition and notation for Olliver-Ricci curvature, the scalar curvature on a graph may be defined to be
\begin{equation} \label{SC} S(x) := \sum_{y}\kappa(x,y).\end{equation}

Note that the scalar curvature on a Riemannian manifold is essentially the contraction of the Ricci curvature tensor with respect to the metric \cite{DoCarmo} (up to scaling factor).  If one only takes adjacent vertices with respect to the hop metric, (\ref{SC}) is compatible with the usual contraction. One may also contract with respect to the family of probability measures, that is, we take the scalar curvature to be defined as
\begin{equation}\label{SC_mod} \hat{S}(x) :=\sum_y \kappa(x,y)\mu_x(y).\end{equation} In practice, both definitions lead to similar results on the networks tested, and we will compare them below.

 Both $S(x)$ and $\hat{S}(x)$ will be seen to provide useful ``nodal measures'' on the set of vertices in the graph, which will become important in comparing previously used entropy measures for biological networks.  For example, the authors in \cite{West} define a notion of entropy (just using direct paths over a graph) as \begin{equation} \label{def:entropy} \bar{S}(x) = - \frac{1}{\log \hat{d}_x} \sum_y \mu_x(y)\log\mu_x(y) \end{equation}
and where $\hat{d}_x$ is the unweighted degree of node $x$.  See \cite{West} for more general definitions when the vertices need not be adjacent.

We will now show in Section \ref{sec:cancer.net} that all forms of curvature presented here indicate a particular ``hallmark'' in cancer networks studied herein.  Finally, we note whereas the Ollivier-Ricci curvature maybe useful in deriving a flow to weaken (or strengthen) a network's robustness, scalar curvature may be useful for identifying nodes (genes) as potential targets and examining the network in a more global fashion. For a nodal measure, it seems that $\hat{S}$ and $\bar{S}$ give comparable results, but the curvature based measure $\hat{S}$ has the numerical advantage of not involving a logarithm that is discontinuous at 0. The Ricci curvature gives a new way of looking at nodal interactions.

\section{Robustness of Cancer Networks} \label{sec:cancer.net}

Of greatest interest to us is the study of cancer networks and the relation of curvature. In particular, we present preliminary results demonstrating that cancer networks exhibit a greater degree of functional robustness with respect to normal tissue networks (as argued by \cite{West}) on both metabolic and transcriptional networks.  We further show that this ``hallmark,'' while significant at the macroscopic network scale, can be utilized to illuminate specific gene-to-gene pairs contributing to the networks robustness (or fragility) with possible associations regarding the onset and proliferation of cancer \cite{ACO1,GOT2,LPO}.  Lastly, we note that quantification by curvature is different than previous quantification methods such as differential gene expression. In what follows, we present a description of the preliminary data and corresponding tabulated results.

\subsection{Brief Description of Data Sets} Our preliminary data (provided by Memorial Sloan Kettering) consists of metabolic and transcriptional regulation networks.  In studying metabolic networks, we are provided eight different cases for which correlation values of gene-to-gene expression in cancer and normal networks are given.  With respect to transcription networks, we are provided two cases of gene-to-gene correlation data sets associated with breast cancer and kidney renal clear cell carcinoma. The network is constructed using these correlation values as weights of the graph and the topology (adjacency matrix) of graph is given by underlying biological gene-to-gene interactions.

\subsection{Metabolic Networks}
The metabolic network data is similar to data utilized in \cite{Ed} with an investigation of an ``arbitrary'' set of 800 genes and their interactions.    An important note regarding this metabolic data is that certain key genes associated with cancer (e.g., BRCA1, TP53) are not present.

\textit{B.1 Metabolic Networks:  Curvature as a Measure for Robustness}\\
Given the above, we first present the difference (cancer-normal) in average curvature of each metabolic network.  This is summarized in Tables \ref{table:all_8_networks_scalar}, \ref{table:all_8_networks}, and \ref{table:all_8_networks_weight} for scalar curvature, Ricci curvature, and weighted scalar curvature, respectively.  We note that the computation here is taken over just direct paths (for necessary comparison to entropy) and that a complete computation (over all possible paths) still yields comparable results presented below. Indeed, since a graph is a 1-geodesic space, if $\kappa(x,y) \ge k$ on adjacent vertices, then this will be true as well for non-adjacent vertices \cite{Ollivier,Oll_markov}. Further, we note that while the average curvature is not the optimal statistic, it nevertheless illustrates that the cancer network as a whole exhibits greater robustness (in terms of curvature) than its corresponding normal tissue network.
\begin{table}[!h]
\mbox{\centering
\parbox{1\columnwidth}{\centering
\begin{tabular}{|c|c|c|c|c|}
\hline
\textbf{Cancer Type} & \textbf{$\Delta$ Scalar Curvature} & \textbf{Cancer Tissue} &  \textbf{Normal Tissue}\\
\hline
\begin{tabular}{c}  Breast Cancer \end{tabular} & \begin{tabular}{c} 0.4284 \end{tabular} & \begin{tabular}{c} 1.9649 \end{tabular} & \begin{tabular}{c}1.5185\end{tabular}\\
\hline
\begin{tabular}{c}  Head / Neck Squamous Cell Carcinoma \end{tabular} & \begin{tabular}{c} 0.3928 \end{tabular} & \begin{tabular}{c} 1.9610 \end{tabular} & \begin{tabular}{c}1.5232\end{tabular}\\
\hline
\begin{tabular}{c}  Kidney Renal Clear Cell Carcinoma \end{tabular} & \begin{tabular}{c} 0.4627 \end{tabular} & \begin{tabular}{c} 1.9724 \end{tabular} & \begin{tabular}{c}1.4647\end{tabular}\\
\hline
\begin{tabular}{c}  Kidney Papillary Carcinoma \end{tabular} & \begin{tabular}{c}  0.5445 \end{tabular} & \begin{tabular}{c} 1.8425 \end{tabular} & \begin{tabular}{c}1.2980\end{tabular}\\
\hline
\begin{tabular}{c}  Liver Cancer  \end{tabular} & \begin{tabular}{c}  0.1088 \end{tabular} & \begin{tabular}{c} 1.8405 \end{tabular} & \begin{tabular}{c}1.7318\end{tabular}\\
\hline
\begin{tabular}{c}  Lung Adenocarcinoma \end{tabular} & \begin{tabular}{c} 0.2059  \end{tabular} & \begin{tabular}{c} 2.0563 \end{tabular} & \begin{tabular}{c}1.8504\end{tabular}\\
\hline
\begin{tabular}{c}  Prostate Cancer \end{tabular} & \begin{tabular}{c}  0.2896 \end{tabular} & \begin{tabular}{c} 1.8970 \end{tabular} & \begin{tabular}{c}1.6074\end{tabular}\\
\hline
\begin{tabular}{c}  Thyroid Cancer  \end{tabular} & \begin{tabular}{c}  0.0829\end{tabular} & \begin{tabular}{c} 1.6937 \end{tabular} & \begin{tabular}{c}1.6108\end{tabular}\\
\hline
\end{tabular}
}}
\caption{Tabulated results for all eight networks with respect to average scalar curvature.  Cancer can be seen to more robust than its complementary normal network in each of the eight cases.}
\label{table:all_8_networks_scalar}
\end{table}
\\

One can see that scalar curvature provides a strong and clear indicator regarding cancer robustness compared to that of healthy tissue.  This being said, in the next two sections, we provide a list of the top twenty gene-to-gene pairs for two of the above cases (given the scope of discussion) to further highlight the proposed cancer ``hallmark.''  Note, unlike that of previous approaches utilizing entropy \cite{West}, we are able to uncover important pairwise information through Ollivier-Ricci curvature.
\begin{table}[!h]
\mbox{\centering
\parbox{1\columnwidth}{\centering
\begin{tabular}{|c|c|c|c|c|}
\hline
\textbf{Cancer Type} & \textbf{$\Delta$ Ricci Curvature} & \textbf{Cancer Tissue} &  \textbf{Normal Tissue}\\
\hline
\begin{tabular}{c}  Breast Cancer \end{tabular} & \begin{tabular}{c} 0.0235 \end{tabular} & \begin{tabular}{c} 0.1071 \end{tabular} & \begin{tabular}{c}0.0836\end{tabular}\\
\hline
\begin{tabular}{c}  Head/Neck Squamous Cell Carcinoma \end{tabular} & \begin{tabular}{c} 0.0217 \end{tabular} & \begin{tabular}{c} 0.1058 \end{tabular} & \begin{tabular}{c}0.0841\end{tabular}\\
\hline
\begin{tabular}{c}  Kidney Renal Clear Cell Carcinoma \end{tabular} & \begin{tabular}{c} 0.0253 \end{tabular} & \begin{tabular}{c} 0.1054 \end{tabular} & \begin{tabular}{c}0.0801\end{tabular}\\
\hline
\begin{tabular}{c}  Kidney Papillary Carcinoma \end{tabular} & \begin{tabular}{c}  0.0300 \end{tabular} & \begin{tabular}{c} 0.1020 \end{tabular} & \begin{tabular}{c}0.0719\end{tabular}\\
\hline
\begin{tabular}{c}  Liver Cancer  \end{tabular} & \begin{tabular}{c}  0.0060 \end{tabular} & \begin{tabular}{c} 0.1011 \end{tabular} & \begin{tabular}{c}0.0951\end{tabular}\\
\hline
\begin{tabular}{c}  Lung Adenocarcinoma \end{tabular} & \begin{tabular}{c} 0.0114  \end{tabular} & \begin{tabular}{c} 0.1137 \end{tabular} & \begin{tabular}{c}0.1023\end{tabular}\\
\hline
\begin{tabular}{c}  Prostate Cancer \end{tabular} & \begin{tabular}{c}  0.0160 \end{tabular} & \begin{tabular}{c} 0.1049 \end{tabular} & \begin{tabular}{c}0.0889\end{tabular}\\
\hline
\begin{tabular}{c}  Thyroid Cancer  \end{tabular} & \begin{tabular}{c}  0.0046\end{tabular} & \begin{tabular}{c} 0.0940 \end{tabular} & \begin{tabular}{c}0.0894\end{tabular}\\
\hline
\end{tabular}
}}
\caption{Tabulated results for all eight networks with respect to average Ricci curvature.  Cancer can be seen to more robust than its complementary normal network in each of the eight cases.}
\label{table:all_8_networks}
\end{table}
\\
\begin{table}[!h]
\mbox{\centering
\parbox{1\columnwidth}{\centering
\begin{tabular}{|c|c|c|c|c|}
\hline
\textbf{Cancer Type} & \textbf{$\Delta$ Weighted Curvature} & \textbf{Cancer Tissue} &  \textbf{Normal Tissue}\\
\hline
\begin{tabular}{c}  Breast Cancer \end{tabular} & \begin{tabular}{c} 0.0139 \end{tabular} & \begin{tabular}{c} 0.1496 \end{tabular} & \begin{tabular}{c}0.1358\end{tabular}\\
\hline
\begin{tabular}{c}  Head / Neck Squamous Cell Carcinoma \end{tabular} & \begin{tabular}{c} 0.0144 \end{tabular} & \begin{tabular}{c} 0.1505 \end{tabular} & \begin{tabular}{c}0.1361\end{tabular}\\
\hline
\begin{tabular}{c}  Kidney Renal Clear Cell Carcinoma \end{tabular} & \begin{tabular}{c} 0.0105 \end{tabular} & \begin{tabular}{c} 0.1515 \end{tabular} & \begin{tabular}{c}0.1410\end{tabular}\\
\hline
\begin{tabular}{c}  Kidney Papillary Carcinoma \end{tabular} & \begin{tabular}{c}  0.0122 \end{tabular} & \begin{tabular}{c} 0.1501 \end{tabular} & \begin{tabular}{c}0.1379\end{tabular}\\
\hline
\begin{tabular}{c}  Liver Cancer  \end{tabular} & \begin{tabular}{c}  0.0054 \end{tabular} & \begin{tabular}{c} 0.1527 \end{tabular} & \begin{tabular}{c}0.1473\end{tabular}\\
\hline
\begin{tabular}{c}  Lung Adenocarcinoma \end{tabular} & \begin{tabular}{c} 0.0087  \end{tabular} & \begin{tabular}{c} 0.1587 \end{tabular} & \begin{tabular}{c}0.1500\end{tabular}\\
\hline
\begin{tabular}{c}  Prostate Cancer \end{tabular} & \begin{tabular}{c}  0.0115 \end{tabular} & \begin{tabular}{c} 0.1495 \end{tabular} & \begin{tabular}{c}0.1380\end{tabular}\\
\hline
\begin{tabular}{c}  Thyroid Cancer  \end{tabular} & \begin{tabular}{c}  0.0069\end{tabular} & \begin{tabular}{c} 0.1430 \end{tabular} & \begin{tabular}{c}0.1361\end{tabular}\\
\hline
\end{tabular}
}}
\caption{Tabulated results for all eight networks with respect to average weighted scalar curvature.  Cancer can be seen to more robust than its complementary normal network in each of the eight cases.}
\label{table:all_8_networks_weight}
\end{table}

\textit{B.2 Metabolic Networks:  Top 20 Gene Pairs of Breast Cancer} \\
In this section, we present the top twenty gene-to-gene pairs (as sorted by difference in Ollivier-Ricci curvature) for the case of breast cancer.  This can be seen in Table \ref{table:Breast_Cancer}.  Most importantly, we should note that by ranking such pairs via differences in curvature, we are able to ``uncover'' specific cancer related genes.  For example, the gene LPO (Lactoperoxidase) has been known to contribute to the initiation of breast cancer \cite{LPO}, SOD3 has been considered an important gene in the defense against oxidative stress and prevention of estrogen-mediated breast cancer \cite{SOD3}, PPAP2C has been described as anti-cancer drug target \cite{PPAP2C},  GOT2 has been noted to significantly affect cell growth of pancreatic ductal adenocarcinoma \cite{GOT2}, and over-expression of LRAT has lead to a poor prognoses in colorectal cancer.  Interestingly, the reoccurrence of genes ACO1 and ACO2 can also be seen in Table \ref{table:Breast_Cancer}.  One possible explanation can be attributed to the fact that these genes regulate iron storage in a given cellular environment, which has recently been highlighted to be connected to increase risk for cancer (in terms of cell proliferation and growth) \cite{ACO1}.  Intuitively, and as argued by \cite{West}, metastasizing cancer exhibits a greater degree of robustness than benign tumors, which in turn, could explain the reasoning of why ACO1 and ACO2 are a common occurrence.

\textit{B.3 Metabolic Networks: Top 20 Gene Pairs of Lung Adenocarcinoma} \\
In this section, we again present the top twenty gene-to-gene pairs with respect to Ricci curvature, but for the case of lung adenocarcinoma.  This can be seen in Table~\ref{table:Lung_Cancer}.  In addition to the biomarkers for lung cancer such as PGDS \cite{PDGS}, it is interesting to note the reoccurrence of the gene UROD (Uroporphyrinogen Decarboxylase).  It has been documented that this gene can be targeted to be ``knocked out'' for a variety of cancers and is a driving force behind several drug therapy methodologies \cite{UROD1,UROD2}.

\begin{table}
\parbox{.45\linewidth}{
\centering
\begin{tabular}{|c|c|c|c|c|}
\hline
Gene & $\Delta$ Ricci Curvature & Gene X &  Gene Y\\ Rank &(Cancer-Normal) &(Symbol) &(Symbol)\\
\hline
\begin{tabular}{c}  1 \end{tabular} & \begin{tabular}{c} 0.4642 \end{tabular} & \begin{tabular}{c} ACAA2 \end{tabular} & \begin{tabular}{c}MYO5B\end{tabular}\\
\hline
\begin{tabular}{c}  2 \end{tabular} & \begin{tabular}{c} 0.4170 \end{tabular} & \begin{tabular}{c} LPO \end{tabular} & \begin{tabular}{c}SOD3\end{tabular}\\
\hline
\begin{tabular}{c}  3 \end{tabular} & \begin{tabular}{c} 0.4162 \end{tabular} & \begin{tabular}{c} BPNT1 \end{tabular} & \begin{tabular}{c}GLYCTK\end{tabular}\\
\hline
\begin{tabular}{c}  4 \end{tabular} & \begin{tabular}{c} 0.4067 \end{tabular} & \begin{tabular}{c} GLYCTK \end{tabular} & \begin{tabular}{c}GAL3ST1\end{tabular}\\
\hline
\begin{tabular}{c}  5 \end{tabular} & \begin{tabular}{c} 0.4057 \end{tabular} & \begin{tabular}{c} PPAP2A \end{tabular} & \begin{tabular}{c}PPAP2C\end{tabular}\\
\hline
\begin{tabular}{c}  6 \end{tabular} & \begin{tabular}{c} 0.3916 \end{tabular} & \begin{tabular}{c} ACO1 \end{tabular} & \begin{tabular}{c}OAT\end{tabular}\\
\hline
\begin{tabular}{c}  7 \end{tabular} & \begin{tabular}{c} 0.3881 \end{tabular} & \begin{tabular}{c} HADHB \end{tabular} & \begin{tabular}{c}MYO5B\end{tabular}\\
\hline
\begin{tabular}{c}  8  \end{tabular} & \begin{tabular}{c} 0.3805 \end{tabular} & \begin{tabular}{c} GUCY2D \end{tabular} & \begin{tabular}{c}NPR2\end{tabular}\\
\hline
\begin{tabular}{c}  9 \end{tabular} & \begin{tabular}{c} 0.3687 \end{tabular} & \begin{tabular}{c} GUCY2D \end{tabular} & \begin{tabular}{c}NPR1\end{tabular}\\
\hline
\begin{tabular}{c}  10 \end{tabular} & \begin{tabular}{c} 0.3624 \end{tabular} & \begin{tabular}{c} GOT2 \end{tabular} & \begin{tabular}{c}ACO2\end{tabular}\\
\hline
\begin{tabular}{c}  11 \end{tabular} & \begin{tabular}{c} 0.3605 \end{tabular} & \begin{tabular}{c} GUCY1A2 \end{tabular} & \begin{tabular}{c}GUCY2D\end{tabular}\\
\hline
\begin{tabular}{c}  12 \end{tabular} & \begin{tabular}{c} 0.3602 \end{tabular} & \begin{tabular}{c} GOT2 \end{tabular} & \begin{tabular}{c}ACO1\end{tabular}\\
\hline
\begin{tabular}{c}  13 \end{tabular} & \begin{tabular}{c} 0.3600 \end{tabular} & \begin{tabular}{c} SGMS1 \end{tabular} & \begin{tabular}{c}LRAT\end{tabular}\\
\hline
\begin{tabular}{c}  14 \end{tabular} & \begin{tabular}{c} 0.3598 \end{tabular} & \begin{tabular}{c} SLC25A22 \end{tabular} & \begin{tabular}{c}SLC25A18\end{tabular}\\
\hline
\begin{tabular}{c}  15 \end{tabular} & \begin{tabular}{c} 0.3584 \end{tabular} & \begin{tabular}{c} AQP9 \end{tabular} & \begin{tabular}{c}AGMAT\end{tabular}\\
\hline
\begin{tabular}{c}  16 \end{tabular} & \begin{tabular}{c} 0.3520 \end{tabular} & \begin{tabular}{c} ETFDH \end{tabular} & \begin{tabular}{c}PYCR1\end{tabular}\\
\hline
\begin{tabular}{c}  17 \end{tabular} & \begin{tabular}{c} 0.3518 \end{tabular} & \begin{tabular}{c} OAT \end{tabular} & \begin{tabular}{c}ACO2\end{tabular}\\
\hline
\begin{tabular}{c}  18 \end{tabular} & \begin{tabular}{c} 0.3449 \end{tabular} & \begin{tabular}{c} GPX3 \end{tabular} & \begin{tabular}{c}LPO\end{tabular}\\
\hline
\begin{tabular}{c}  19 \end{tabular} & \begin{tabular}{c} 0.3373 \end{tabular} & \begin{tabular}{c} ALDH9A1 \end{tabular} & \begin{tabular}{c}ALDH7A1\end{tabular}\\
\hline
\begin{tabular}{c}  20 \end{tabular} & \begin{tabular}{c} 0.3317 \end{tabular} & \begin{tabular}{c} PYGB \end{tabular} & \begin{tabular}{c}PYGL\end{tabular}\\
\hline
\end{tabular}
\caption{The top 20 gene pairs in the metabolic network for breast tissue case sorted with respect to Ricci curvature.}
\label{table:Breast_Cancer}
}
\hfill
\parbox{.45\linewidth}{
\centering
\begin{tabular}{|c|c|c|c|c|}
\hline
Gene & $\Delta$ Ricci Curvature & Gene X &  Gene Y\\ Rank &(Cancer-Normal) &(Symbol) &(Symbol)\\
\hline
\begin{tabular}{c}  1 \end{tabular} & \begin{tabular}{c} 0.3418 \end{tabular} & \begin{tabular}{c} PGDS \end{tabular} & \begin{tabular}{c}SLCO2A1\end{tabular}\\
\hline
\begin{tabular}{c}  2 \end{tabular} & \begin{tabular}{c} 0.2807 \end{tabular} & \begin{tabular}{c} NEU4 \end{tabular} & \begin{tabular}{c}CMAS\end{tabular}\\
\hline
\begin{tabular}{c}  3 \end{tabular} & \begin{tabular}{c} 0.2364 \end{tabular} & \begin{tabular}{c} HGD \end{tabular} & \begin{tabular}{c}HPD\end{tabular}\\
\hline
\begin{tabular}{c}  4 \end{tabular} & \begin{tabular}{c} 0.2322 \end{tabular} & \begin{tabular}{c} SLC26A2 \end{tabular} & \begin{tabular}{c}UROD\end{tabular}\\
\hline
\begin{tabular}{c}  5 \end{tabular} & \begin{tabular}{c} 0.2287 \end{tabular} & \begin{tabular}{c} SLC26A3 \end{tabular} & \begin{tabular}{c}UROD\end{tabular}\\
\hline
\begin{tabular}{c}  6 \end{tabular} & \begin{tabular}{c} 0.2201 \end{tabular} & \begin{tabular}{c} PRDX3 \end{tabular} & \begin{tabular}{c}CYP24A1\end{tabular}\\
\hline
\begin{tabular}{c}  7 \end{tabular} & \begin{tabular}{c} 0.2177 \end{tabular} & \begin{tabular}{c} SLC26A1 \end{tabular} & \begin{tabular}{c}UROD\end{tabular}\\
\hline
\begin{tabular}{c}  8  \end{tabular} & \begin{tabular}{c} 0.2085 \end{tabular} & \begin{tabular}{c} PPAP2A \end{tabular} & \begin{tabular}{c}DGAT1\end{tabular}\\
\hline
\begin{tabular}{c}  9 \end{tabular} & \begin{tabular}{c} 0.2011 \end{tabular} & \begin{tabular}{c} GLYCTK \end{tabular} & \begin{tabular}{c}GAL3ST1\end{tabular}\\
\hline
\begin{tabular}{c}  10 \end{tabular} & \begin{tabular}{c} 0.1976 \end{tabular} & \begin{tabular}{c} SLC26A11 \end{tabular} & \begin{tabular}{c}UROD\end{tabular}\\
\hline
\begin{tabular}{c}  11 \end{tabular} & \begin{tabular}{c} 0.1967 \end{tabular} & \begin{tabular}{c} SGMS1 \end{tabular} & \begin{tabular}{c}DGAT1\end{tabular}\\
\hline
\begin{tabular}{c}  12 \end{tabular} & \begin{tabular}{c} 0.1924 \end{tabular} & \begin{tabular}{c} SLC26A8 \end{tabular} & \begin{tabular}{c}UROD\end{tabular}\\
\hline
\begin{tabular}{c}  13 \end{tabular} & \begin{tabular}{c} 0.1860 \end{tabular} & \begin{tabular}{c} NEU4 \end{tabular} & \begin{tabular}{c}SLC17A5\end{tabular}\\
\hline
\begin{tabular}{c}  14 \end{tabular} & \begin{tabular}{c} 0.1849 \end{tabular} & \begin{tabular}{c} ACADS \end{tabular} & \begin{tabular}{c}ACAD11\end{tabular}\\
\hline
\begin{tabular}{c}  15 \end{tabular} & \begin{tabular}{c} 0.1826 \end{tabular} & \begin{tabular}{c} SLC26A6 \end{tabular} & \begin{tabular}{c}UROD\end{tabular}\\
\hline
\begin{tabular}{c}  16 \end{tabular} & \begin{tabular}{c} 0.1817 \end{tabular} & \begin{tabular}{c} CHIA \end{tabular} & \begin{tabular}{c}RENBP\end{tabular}\\
\hline
\begin{tabular}{c}  17 \end{tabular} & \begin{tabular}{c} 0.1806 \end{tabular} & \begin{tabular}{c} SLC26A9 \end{tabular} & \begin{tabular}{c}UROD\end{tabular}\\
\hline
\begin{tabular}{c}  18 \end{tabular} & \begin{tabular}{c} 0.1762 \end{tabular} & \begin{tabular}{c} CPT1B \end{tabular} & \begin{tabular}{c}PIGW\end{tabular}\\
\hline
\begin{tabular}{c}  19 \end{tabular} & \begin{tabular}{c} 0.1737 \end{tabular} & \begin{tabular}{c} CPT1A \end{tabular} & \begin{tabular}{c}PIGW\end{tabular}\\
\hline
\begin{tabular}{c}  20 \end{tabular} & \begin{tabular}{c} 0.1727 \end{tabular} & \begin{tabular}{c} HYAL3 \end{tabular} & \begin{tabular}{c}LPGAT1\end{tabular}\\
\hline
\end{tabular}
\caption{The top 20 gene pairs in the metabolic network for lung adenocarcinoma tissue case sorted with respect to Ricci curvature.}
\label{table:Lung_Cancer}
}
\end{table}

\textit{B.4 Metabolic Networks:  Comparison of Entropy and Curvature} \\
In the previous sections, we utilized the fact that curvature is positively correlated with that of entropy to quantify the robustness of a network.  While these two metrics may be used, the discriminatory power of each measure may differ.  In this section, we provide a comparison of entropy with that of scalar and Ricci curvature (and hence repeat some results for the sake of clarity).  Utilizing equation (\ref{def:entropy}), we are able to compute entropy as a nodal measure and fairly compare that to that of scalar curvature (also a nodal measure).  For completeness, we repeat the average difference in Ricci curvature.  These results are provided in Table \ref{table:all_8_networks_diff}.
\begin{table}[!h]
\mbox{\centering
\parbox{1\columnwidth}{\centering
\begin{tabular}{|c|c|c|c|c|}
\hline
\textbf{Cancer Type} & \textbf{$\Delta$ Entropy} & \textbf{$\Delta$ Scalar Curvature} &  \textbf{$\Delta$ Ricci Curvature}\\
\hline
\begin{tabular}{c}  Breast Cancer \end{tabular} & \begin{tabular}{c} 0.0146 \end{tabular} & \begin{tabular}{c} 0.4284 \end{tabular} & \begin{tabular}{c}0.0234\end{tabular}\\
\hline
\begin{tabular}{c}  Head / Neck Squamous Cell Carcinoma \end{tabular} & \begin{tabular}{c} 0.0152 \end{tabular} & \begin{tabular}{c} 0.3928 \end{tabular} & \begin{tabular}{c}0.0216\end{tabular}\\
\hline
\begin{tabular}{c}  Kidney Renal Clear Cell Carcinoma \end{tabular} & \begin{tabular}{c} 0.0168 \end{tabular} & \begin{tabular}{c} 0.4627 \end{tabular} & \begin{tabular}{c}0.0252\end{tabular}\\
\hline
\begin{tabular}{c}  Kidney Papillary Carcinoma \end{tabular} & \begin{tabular}{c}  0.0189 \end{tabular} & \begin{tabular}{c} 0.5455 \end{tabular} & \begin{tabular}{c}0.0300\end{tabular}\\
\hline
\begin{tabular}{c}  Liver Cancer  \end{tabular} & \begin{tabular}{c}  0.0030 \end{tabular} & \begin{tabular}{c} 0.1088 \end{tabular} & \begin{tabular}{c}0.0060\end{tabular}\\
\hline
\begin{tabular}{c}  Lung Adenocarcinoma \end{tabular} & \begin{tabular}{c} 0.0049  \end{tabular} & \begin{tabular}{c} 0.2059 \end{tabular} & \begin{tabular}{c}0.0113\end{tabular}\\
\hline
\begin{tabular}{c}  Prostate Cancer \end{tabular} & \begin{tabular}{c}  0.0110 \end{tabular} & \begin{tabular}{c} 0.2896 \end{tabular} & \begin{tabular}{c}0.0159\end{tabular}\\
\hline
\begin{tabular}{c}  Thyroid Cancer  \end{tabular} & \begin{tabular}{c}  0.0031\end{tabular} & \begin{tabular}{c} 0.0829 \end{tabular} & \begin{tabular}{c}0.0046\end{tabular}\\
\hline
\end{tabular}
}}
\caption{Comparison of Entropy and Different Notions of Curvature.}
\label{table:all_8_networks_diff}
\end{table}

We again note that from a computationally comparison standpoint, the proposed usage of a curvature can be simply implemented as a highly parallelizable linear program compared to that of \cite{West}.  Thus, for paths that are larger than one, our computation remains  unaffected.  Moreover, entropy defined in equation (\ref{def:entropy}) requires a minimum of two incidences per node in order for the measure to be well-behaved. Utilizing Ricci (or scalar) curvature, we do not impose such restrictions.   This is particularly important, as many networks possess a degree distribution that follows the power law \cite{barabasi}.

\subsection{Transcriptional Networks}
The transcription network data consists of two of the above cancer cases with investigation of 450 cancer-related genes and their interactions.    An important note, unlike that of the metabolic data, we focus on the inter-relationship of cancer related genes (e.g., BRCA1, TP53 are present).

\textit{C.1 Transcriptional Networks:  Curvature as a Measure for Robustness}\\
Similar to that of B.1 and for the sake of brevity, this section presents the average difference (cancer-normal) in scalar curvature of two transcription networks provided by MSKCC.  In Table \ref{table:all_2_networks_scalar}, we again note that scalar curvature is a clear indicator of robustness and while not shown below, average Ricci curvature also yields similar results.
\begin{table}[!h]
\mbox{\centering
\parbox{1\columnwidth}{\centering
\begin{tabular}{|c|c|c|c|c|}
\hline
\textbf{Cancer Type} & \textbf{$\Delta$ Scalar Curvature} & \textbf{Cancer Tissue} &  \textbf{Normal Tissue}\\
\hline
\begin{tabular}{c}  Breast Cancer \end{tabular} & \begin{tabular}{c} 0.4292 \end{tabular} & \begin{tabular}{c} 8.2037 \end{tabular} & \begin{tabular}{c}7.7744\end{tabular}\\
\hline
\begin{tabular}{c}  Kidney Renal Clear Cell Carcinoma \end{tabular} & \begin{tabular}{c} 0.0728 \end{tabular} & \begin{tabular}{c} 8.1460 \end{tabular} & \begin{tabular}{c}8.7033\end{tabular}\\
\hline
\end{tabular}
}}
\caption{Tabulated results for the two transcription networks with respect to average scalar curvature.  Cancer can be seen to more robust than its complementary normal network in each of the two cases.}
\label{table:all_2_networks_scalar}
\end{table}

\textit{C.2 Transcriptional Networks: Top and Bottom 10 Pairs}\\
It is of particular interest to examine a network at the local level to illuminate certain attributes of cancer.  In this section, we present the top and bottom ten gene-to-gene pairs of the possible interactions  (as sorted by difference in Ollivier-Ricci curvature) for the transcription network cases of breast cancer and kidney renal clear cell carcinoma.  This can be seen in Table \ref{table:top10_Breast}, Table \ref{table:top10_Kidney}, Table \ref{table:bot10_Breast}, and Table \ref{table:bot10_Kidney}

From the aforementioned tables, a few insights are noted.  Firstly and in the particular case of breast cancer, RNF43 and TP53 are known tumor suppressor genes and ranked in the top and last gene pairs (out of a 100K possible pairs).  In fact, from Table  \ref{table:top10_Breast} and Table \ref{table:bot10_Kidney}, one can see several genes related to cell proliferation, division, and growth (e.g., GPC3, POT1, RNF43, TP53).  With respect to kidney renal clear cell carcinoma, it is interesting to note the reoccurrence of PAX8 and its role in adult kidney and renal tumors.  Such large differences relating to an increase/decrease in robustness for such genes require further analysis and provides an exciting avenue into understanding the overall robustness of cancer networks.

Additionally, one more important observation regarding the above computation is the uncovering of ``hidden'' pairs (denoted by * in gene ranking).  That is, we compute curvature here on all possible pair/pathways as opposed to pairs only defined by the underlying biological adjacency matrix and results given above. Examining all possible pairs (in a feasible linear program computation) further illuminates possible important inter-relationships in a given network (e.g., interactions with PAX8).
\begin{table}
\parbox{.45\linewidth}{
\centering
\begin{tabular}{|c|c|c|c|}
\hline
Gene &   $\Delta$ Ricci Curvature & Gene X &  {Gene Y}\\ Ranking &(Cancer-Normal) &(Symbol) &(Symbol)\\
\hline
1&   0.3504   &  RNF43  		&RSPO3\\
\hline
2&  0.3444    &   RNF43    	&RSPO2\\
\hline
3&  0.3012    &  ERG 		&ETV1\\
\hline
4&  0.3001    &  GPC3 		&PTCH1\\
\hline
5&  0.2901    &  GPC3		 &SDC4\\
\hline
6&  0.2796    &  POT1		 &SBDS\\
\hline
7& 0.2538     &   FGFR2 		&KDR\\
\hline
8*& 0.2510    &  ETV1  		 &FEV\\
\hline
9&  0.2460    &   ERG     		&FOXA1\\
\hline
10&  0.2410 &    EXT1   		 &SDC4\\
\hline
\end{tabular}
\caption{Top 10 Pairs in Breast Cancer. }
\label{table:top10_Breast}
}
\hfill
\parbox{.45\linewidth}{
\centering
\begin{tabular}{|c|c|c|c|}
\hline
Gene &   $\Delta$ Ricci Curvature & Gene X &  Gene Y\\ Ranking &(Cancer-Normal) &(Symbol) &(Symbol)\\
\hline
1&   0.2370   &  LMO1  		&LYL1\\
\hline
2&   0.2300   &  LMO1  		&TAL1\\
\hline
3&   0.2204   &  MAF  		&NFATC2\\
\hline
4&   0.1837   &  DDX10  		&NUP214\\
\hline
5*&   0.1827   &  DDX10  		&ELN\\
\hline
6&   0.1688   &  ERBB2  		&PDGFRB\\
\hline
7&   0.1681   &  IDH1  		&SBDS\\
\hline
8&   0.1661   &  FLT3  		&ROS1\\
\hline
9&   0.1613   &  NTRK3  		&ROS1\\
\hline
10&   0.1601   &  IL6ST  		&MITF\\
\hline
\end{tabular}
\caption{Top 10 Pairs in Kidney Cancer. }
\label{table:top10_Kidney}
}
\end{table}

\begin{table}
\parbox{.45\linewidth}{
\centering
\begin{tabular}{|c|c|c|c|}
\hline
Gene &   $\Delta$ Ricci Curvature & Gene X &  Gene Y\\ Ranking &(Cancer-Normal) &(Symbol) &(Symbol)\\
\hline
99226*&   -0.2203   &  POT1  		&RNF43\\
\hline
99227&   -0.2214   &  ELF4  		&FEV\\
\hline
99228*&   -0.2304   &  MUC1  		&RNF43\\
\hline
99229&   -0.2489   &  IDH1  		&SBDS\\
\hline
99230*&   -0.2496   &  ERCC2  		&RNF43\\
\hline
99231*&   -0.2839   &  PRDM1  		&RNF43\\
\hline
99232&   -0.2851   &  CCND3  		&ETV1\\
\hline
99233&   -0.3632   &  RNF43  		&SFPQ\\
\hline
99234&   -0.3636   &  NONO  		&RNF43\\
\hline
99235&   -0.3651   &  RNF43  		&TP53\\
\hline\end{tabular}
\caption{Bottom 10 Pairs in Breast Cancer.}
\label{table:bot10_Breast}
}
\hfill
\parbox{.45\linewidth}{
\centering
\begin{tabular}{|c|c|c|c|}
\hline
Gene &   $\Delta$ Ricci Curvature & Gene X &  Gene Y\\ Ranking &(Cancer-Normal) &(Symbol) &(Symbol)\\
\hline
99226*&   -0.1730   &  KDM5A  		&PAX8\\
\hline
99227*&   -0.1731   &  BCL11A  		&PAX8\\
\hline
99228*&   -0.1766   &  PAX8  		&ZNF521\\
\hline
99229*&   -0.1821   &  PAX8  		&PWWP2A\\
\hline
99230&   -0.1857   &  POT1  		&SBDS\\
\hline
99231&   -0.1878   &  GPC3  		&PTCH1\\
\hline
99232&   -0.1900   &  CREB3L2        &GPHN\\
\hline
99233&   -0.1903   &  DDX10  		&MYCL1\\
\hline
99234&   -0.1935   &  ELF4  		&FEV\\
\hline
99235&   -0.1942   &  CD74  		&MYC\\
\hline
\end{tabular}
\caption{Bottom 10 Pairs in Kidney Cancer. }
\label{table:bot10_Kidney}
}
\end{table}

\textit{B.4 Transcriptional Networks:  Comparison of Entropy and Curvature} \\
Similar to that of B.4 and for the sake of brevity, this section presents a comparison of entropy with that of curvature for the two transcription networks provided by MSKCC.  In Table \ref{table:all_2_networks_diff}, we compare scalar curvature and entropy as measures of robustness.
\begin{table}[!h]
\mbox{\centering
\parbox{1\columnwidth}{\centering
\begin{tabular}{|c|c|c|c|c|}
\hline
\textbf{Cancer Type} & \textbf{$\Delta$ Entropy} & \textbf{$\Delta$ Scalar Curvature} &  \textbf{$\Delta$ Ricci Curvature}\\
\hline
\begin{tabular}{c}  Breast Cancer \end{tabular} & \begin{tabular}{c} 0.0117 \end{tabular} & \begin{tabular}{c} 0.4292 \end{tabular} & \begin{tabular}{c}0.0117\end{tabular}\\
\hline
\begin{tabular}{c}  Kidney Renal Clear Cell Carcinoma \end{tabular} & \begin{tabular}{c} 0.0060 \end{tabular} & \begin{tabular}{c} 0.0728 \end{tabular} & \begin{tabular}{c}0.0020\end{tabular}\\
\hline
\end{tabular}
}}
\caption{Tabulated results for the two transcription networks with respect to average scalar curvature.  Cancer can be seen to more robust than its complementary normal network in each of the two cases.}
\label{table:all_2_networks_diff}
\end{table}

\section{Conclusion and Future Research}

In this work, we have considered using the geometric notion of curvature in order to study the functional robustness of cancer networks. There are some interesting related ideas that we will be considering in future work that may also have some bearing in network changes in topology.

First, in some very interesting work, Billera {\em et al.} \cite{Billera} describe a metric geometry on the space of trees in connection with phylogenetics. It turns out that their space is a moduli space (universal parameter space) and has non-positive curvature. From the results of \cite{Sturm}, this allows one to do statistics on this space since between any two points there is a unique geodesic. This has had a number of intriguing applications including to cancer research \cite{Rabadan}. It would be very interesting to generalize this to more general network structures, and instead of just looking at the geometric (curvature) property of an individual network to devise quantitative statistical methods based on the metric geometry comparing families of networks.

The work of \cite{Rabadan} is largely motivated by the problem of cancer cell heterogeneity. Indeed, cancer progression is believed to follow Darwinian evolutionary pattern: fitter subtypes replace other less fit cells, which leads to disease. In combination with high-throughput genomics one can construct trees to study this process. This is an example of a deep relationship between the concepts of Darwinian evolution and Boltzmann thermodynamics advocated by Demetrius \cite{demo}. The idea is that macroscopic entropy increases under microscopic molecular collisions, while macroscopic evolution can be (partially) explained via the concept of the increase of entropy. This reasoning is very much in line with the overall thrust of the present paper in which we are trying to use curvature (positively correlated with robustness) to quantify network robustness. The macroscopic theory is also very much in line with Boltzmann thermodynamics. Evolutionary changes and network adaptability are key topics to be considered in future research.

\end{document}